\documentclass[twocolumn,aps,prl,showpacs,amsmath]{revtex4-1}
\usepackage{amsfonts}
\usepackage{amssymb}
\usepackage{mathrsfs}
\usepackage{graphicx}
\usepackage{float}
\usepackage{xcolor}
\usepackage{bm}

\begin{document}

\title{Exact mobility edges in Aubry-Andr\'{e}-Harper models with relative phases}

\author{Xiaoming Cai}
\email{cxmpx@wipm.ac.cn}

\author{Yi-Cong Yu}
\email{ycyu@wipm.ac.cn}

\address{State Key Laboratory of Magnetic Resonance and Atomic and Molecular Physics, Wuhan Institute of Physics and Mathematics, APM, Chinese Academy of Sciences, Wuhan 430071, China}
\date{\today}

\begin{abstract}
Mobility edge (ME), a critical energy separating localized and extended states in spectrum, is a central concept in understanding the localization physics. 
However, there are few models with exact MEs. 
In the paper, we generalize the Aubry-Andr\'{e}-Harper model proposed in [Phys. Rev. Lett. \textbf{114}, 146601 (2015)] and recently realized in [Phys. Rev. Lett. \textbf{126}, 040603 (2021)], by introducing a relative phase in the quasiperiodic potential. 
Applying Avila's global theory we analytically compute localization lengths of all single-particle states and determine the exact expression of ME, which both significantly depend on the relative phase. 
They are verified by numerical simulations, and a physical perception of the exact expression is also provided. 
We further demonstrate that the exact expression of ME works for an even broad class of generalized Aubry-Andr\'{e}-Harper models. 
Moreover, we show that the exact ME is related to the one in the dual model which has long-range hoppings.
\end{abstract}
\maketitle

\section{I. Introduction}

Ever since the seminal work of Anderson, quantum localization has been a fundamental and extensively studied phenomenon in condensed matter physics \cite{Anderson1958,Abrahams2010}, in which the disorder induces localized states and leads to the absence of diffusion. 
According to scaling theory, an infinitesimal random disorder exponentially localizes all single-particle states in one and two dimensions \cite{Evers2008}. 
Whereas localized and extended states can coexist at different energies in three-dimensional Anderson systems. 
There is a critical energy, known as single-particle mobility edge (SPME), and separating localized and extended eigenstates in spectrum \cite{Evers2008,Lagendijk2009}. 
SPME is crucial in understanding various fundamental phenomena, like the metal-insulator transition, and a system with it has a strong thermoelectric response which can be used in thermoelectric devices \cite{Whitney2014,Yamamoto2017,Chiaracane2020}. 
However, experimentally it is still a challenge to determine the SPME in three-dimensional Anderson systems, and theoretically it is hard to introduce microscopic models to understand the physics of SPME in three dimensions, especially the ones with exact SPMEs for analytical studies.
As an intermediate case between disordered and periodic systems, quasiperiodic ones exhibit distinct behaviors, and may support localization phase transitions and SPMEs even in one dimension (1D). 
A well-known example is the 1D Aubry-Andr\'{e}-Harper (AAH) model \cite{Aubry1980,Harper1955}. 
It undergoes a transition from a completely extended phase to a completely localized phase at a finite strength of the quasiperiodic potential, guaranteed by a self-duality. 
It has energy-independent localization lengths of states, and there is no SPME.
%
The AAH model and its various extensions have been theoretically studied extensively \cite{Sarma1998,Cai2013,Rossignolo2019,Biddle2010,Ganeshan2015,Deng2019,Wang2020,Roy2021,Wang2021a,Liu2015}, and experimentally realized in a variety of systems, such as ultracold atoms \cite{Roati2008,Li2017,Luschen2018,Kohlert2019,Goblot2020,An2021,An2018} and photonic crystals \cite{Lahini2009,Kraus2012}.
By introducing short- \cite{Roy2021} or long-range hopping term \cite{Deng2019}, spin-orbit coupling \cite{Zhou2013,Kohmoto2008}, or modified quasiperiodic potentials breaking the self-duality of the classic AAH model \cite{Biddle2009,Li2017,Yao2019,Saha2019}, one can obtain SPMEs in various generalized AAH models. 
However, very few of them support exact expressions of SPMEs \cite{Biddle2010,Ganeshan2015,Wang2020,Liu2021,Bodyfelt2014}. 
Biddle \textit{et al.} found exact SPMEs in a generalized AAH model with long-range hoppings \cite{Biddle2010}, and Ganeshan \textit{et al.} revealed the existence in a model with modified quasiperiodic potential \cite{Ganeshan2015}. 
Two models are related by a duality transformation \cite{Wang2021}, and they are dual to each other. 
Exact SPMEs in them are guaranteed by a self-duality mapping, which is recently applied to a similar model with exactly invariable SPMEs \cite{Liu2021}. 
Other exact SPMEs were found in a family of quasiperiodic mosaic lattices, by analytically computing the localization length \cite{Wang2020}. 
These few models with exact SPMEs provide benchmark understandings of the SPME physics in both noninteracting and interacting systems, and searching for more is highly desirable.
In the paper, we generalize the AAH model proposed by Ganeshan \cite{Ganeshan2015}, by introducing a relative phase in the quasiperiodic potential. 
We analytically compute the Lyapunov exponents (LEs) (or inverse of localization lengths) of all single-particle states, by applying Avila's global theory of one-frequency Schr\"{o}dinger operators \cite{Avila2015}. 
Then, not only an exact expression of SPME is obtained, but also localization properties, which both are significantly changed by the relative phase. 
Numerical verifications of them and a physical perception of the exact expression are given in order. 
Furthermore, we show that the exact expression of SPME works for an even broad class of generalized AAH models. 
Besides, exact SPMEs in dual models, which have long-range hopping terms and opposite localization properties, are also studied.

The rest of paper is organized as follows. 
In section II, we introduce generalized AAH models with relative phases. 
Section III is devoted to the analytical computation of the LEs of all single-particle states. 
Then the exact expression of SPME is obtained from the computed LEs. 
Numerical verifications are done by computing inverse of the participation ratio and the fractal dimension. 
In section IV, we study exact SPMEs in dual models and in an even broad class of generalized AAH models. 
Finally, we present a conclusion and discussion in section V.

\section{II. Model and Hamiltonian}

We consider a family of generalized AAH models, described by the Hamiltonian
\begin{equation}
H=t\sum_j(c^\dagger_jc_{j+1}+c^\dagger_{j+1}c_j)+\sum_jV_jc^\dagger_jc_j.
\label{Ham}
\end{equation}
$c^\dagger_j(c_j)$ is the creation (annihilation) operator of a particle at site $j$. 
$t$ is the hopping amplitude and sets the unit of energy ($t=1$). 
The quasiperiodic potential is given by
\begin{equation}
V_j=2V\frac{\mathrm{cos}(2\pi\beta j+\theta+\delta)}{1-b\mathrm{cos}(2\pi\beta j+\theta)}.
\label{QP}
\end{equation}
Compared to the one studied in Ref.\cite{Ganeshan2015}, a relative phase $\delta$ is introduced here, which changes the SPME and localization properties significantly. 
$V$ is the strength of quasiperiodic potential.  
$\beta$ is an irrational number, characterizing the quasiperiodicity.
It usually takes the value of golden ratio [$\beta=(\sqrt{5}+1)/2$]. 
In practice, it is approximated by rational numbers $\beta=F_{n+1}/F_{n}$ with $F_n$ the $n$th Fibonacci number, and the number of lattice sites $L=F_n$. 
To ensure bounded nature of the quasiperiodic potential, $b\in(-1,1)$ is required, and the potential is a smooth function of $b$. 
Although a restriction on $b$ is made in the paper, the analytical computation presented below works for any $b$, regardless of the possible unbounded nature. 
$\theta$ is a global phase, which is trivial on localization, and we will set $\theta=0$ if not specified. 
In the presence of phases $\theta$ and $\delta$, we can set $V$ positive real and $b\in[0,1)$.
The model reduces to the classic AAH model when $b=0$ \cite{Aubry1980}, which manifests a localization phase transition at the self-duality point $V=t$. 
When $V<t$ the system is in the extended phase where all single-particle states in spectrum are extended, whereas it is in the localized phase when $V>t$ and all states are localized with energy-independent localization lengths. 
There is no SPME.
When $\delta=0$ or $\pi$, the model reduces to the ones studied in Ref. \cite{Ganeshan2015}, where exact expressions of SPMEs $|bE/2\pm V|=t$ were determined by a self-duality mapping. 
In the ($V,E$) plane, states with energies $E$ between above two parallel lines of SPMEs are extended, while outside the lines they are localized.

\section{III. Exact expressions of the LE and SPME}

\begin{figure*}[tbp]
	\begin{center}
		\includegraphics[scale=1, bb=0 0 455 251]{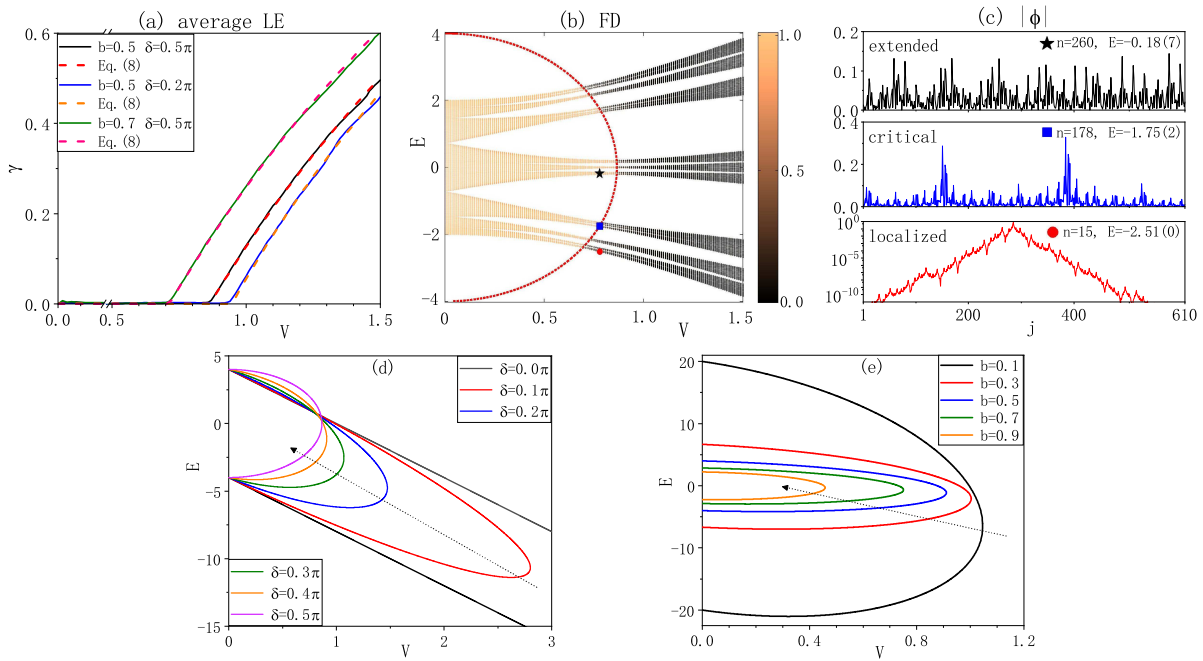}
		\caption{(a) Average LEs vs. $V$, of states with zero energy, for systems with different $b$ and $\delta$. 
			(b) FDs of single-particle states, as function of the corresponding energies $E$ and quasiperiodic potential strength $V$, for systems with $b=0.5$ and $\delta=\pi/2$. 
			Red dash line corresponds to the theoretical prediction in Eq.(\ref{MB}). 
			(c) Typical spacial distributions of states, whose positions in spectra are shown in (b) (symbols).
			$n$ is the index of states which are arranged in the ascending order of energies. 
			Lines of SPMEs are shown in (d) for systems with $b=0.5$ and different $\delta$, and in (e) for systems with $\delta=0.4\pi$ and different $b$. 
			The number of lattice sites $L=610$.}
		\label{Fig1}
	\end{center}
\end{figure*}

The self-duality mapping, proposed in Ref.\cite{Ganeshan2015} to determine exact SPMEs of the model with $\delta=0$ or $\pi$, does not work for the model with a general $\delta$. 
Here, we determine exact expressions of SPMEs by analytically computing LEs of single-particle states. 
Given an eigenstate $|\Phi\rangle=\sum_j\phi_jc^\dagger_j|0\rangle$ with energy $E$, the Schr\"{o}dinger equation of amplitudes $\phi_j$ in transfer matrix form is written as
\begin{equation}
\left[\begin{array}{c}
\phi_{j+1}\\
\phi_j\end{array}
\right]=T_j\left[\begin{array}{c}
\phi_{j}\\
\phi_{j-1}\end{array}
\right],T_j=\left[\begin{array}{cc}
\frac{E-V_j}{t}&-1\\
1&0\end{array}
\right].
\label{TM}
\end{equation}
The LE of state is computed by
\begin{equation}
\gamma(E)=\lim_{L\rightarrow\infty}\frac{1}{L}\mathrm{ln}\|\prod_{j=1}^LT_j\|,
\end{equation}
where $\|\cdot\|$ denotes the norm of a matrix, which is defined by the largest absolute value of its eigenvalues. 
Notice that the transfer matrix can be rewritten as a commutative product
\begin{eqnarray}
T_j=&&\frac{1}{1-b\mathrm{cos}(2\pi\beta j+\theta)}\times\notag\\
&&\left[\begin{array}{cc}
\frac{E[1-b\mathrm{cos}(2\pi\beta j+\theta)]-2V\mathrm{cos}(2\pi\beta j+\delta+\theta)}{t}&b\mathrm{cos}(2\pi\beta j+\theta)-1\\
1-b\mathrm{cos}(2\pi\beta j+\theta)&0\end{array}
\right],\notag\\
=&&A_jB_j.
\end{eqnarray}
Applying ergodic theory and the Jensen's formula, we obtain the LE of part $A_j$
\begin{eqnarray}
\gamma^A(E)&=&\lim_{L\rightarrow\infty}\frac{1}{L}\mathrm{ln}\prod_{j=1}^L\frac{1}{|1-b\mathrm{cos}(2\pi\beta j+\theta)|},\\
&=&\frac{1}{2\pi}\int_0^{2\pi}\mathrm{ln}\frac{1}{|1-b\mathrm{cos}\theta|}d\theta=\mathrm{ln}\frac{2}{1+\sqrt{1-b^2}}.\notag
\end{eqnarray}
As for the LE of part $B_j$, we employ Avila's global theory of one-frequency analytical $SL(2,\mathbb{R})$ cocycle \cite{Avila2015}. 
The first step in the computation is to carry out an analytical continuation of the global phase, i.e., $\theta\rightarrow\theta+i\varepsilon$. 
In the absence of ambiguity, we will use the same symbol for a quantity and its analytical continuation. 
In the limit $\varepsilon\rightarrow\infty$, a straightforward calculation yields
\begin{equation}
	B_j(\varepsilon\rightarrow\infty)=e^{-i2\pi\beta j-i\theta+\varepsilon}\left[\begin{array}{cc}
	-\frac{bE}{2t}-\frac{V}{t}e^{-i\delta}&\frac{b}{2}\\
	-\frac{b}{2}&0\end{array}
	\right]+o(1),
\end{equation}
which results in $\gamma^B_{\varepsilon\rightarrow\infty}(E)=|\varepsilon|+\max\{\ln|(bE+2Ve^{-i\delta})\pm\sqrt{(bE+2Ve^{-i\delta})^2-4b^2t^2}|/4t\}$. 
According to Avila's global theory, as a function of $\varepsilon$, $\gamma^B_\varepsilon(E)$ is a convex, piecewise linear function with integer slopes.
Moreover, the theory shows that, the energy $E$ does not belong to the spectrum, if and only if $\gamma^B_{\varepsilon=0}(E)>0$ and $\gamma^B_\varepsilon(E)$ is an affine function in the neighborhood of $\varepsilon=0$. 
Including $\gamma^A(E)$, we obtain the LE of the single-particle state
\begin{equation}
\gamma(E)=\gamma^A+\gamma^B_{\varepsilon=0}=\max\left(f,0\right),
\label{LE}
\end{equation}
where $f=\max\{\ln|(bE+2Ve^{-i\delta})\pm\sqrt{(bE+2Ve^{-i\delta})^2-4b^2t^2}|/2t(1+\sqrt{1-b^2})\}$. 
Condition $\gamma(E)\geq 0$ has been used above. 
The quantity $f$ and then the LE are periodic functions of the relative phase $\delta$ with a period $2\pi$. 
Moreover, they are symmetric with respect to $\delta=m\pi, m\in\mathbb{Z}$. 
Thus, we can restrict $\delta\in[0,\pi]$. 
And $f$ can be further simplified to
\begin{widetext}
\begin{equation}
f=\ln\left|\frac{|bE+2V\cos\delta|+i2V\sin\delta+\sqrt{(|bE+2V\cos\delta|+i2V\sin\delta)^2-4b^2t^2}}{2t(1+\sqrt{1-b^2})}\right|.
\end{equation}
\end{widetext}
SPMEs are determined by the condition $f=0$, which gives the exact expression
\begin{equation}
\left(\frac{bE}{2}+V\cos\delta\right)^2+\frac{V^2\sin^2\delta}{1-b^2}=t^2.
\label{MB}
\end{equation}
These are the central results of this work.

The analytical results are consistent with numerical simulations. 
In Fig.\ref{Fig1}(a) we present examples of average LEs vs. $V$, of the states with zero energy,  for systems with different $b$ and $\delta$. 
We adopt exponentially decaying wave functions $\phi^n_j=\mathrm{exp}(-\gamma_n|j-j_0|)$ with $j_0$ the localization center, $n$ the index of states, and $\gamma_n$ the LE. 
Then LEs are extracted by fitting numerical single-particle states with above wave functions. 
The averaging is done over states around energy $E=0$, and it is further performed over  $200$ realizations of the quasiperiodic potential with global phases $\theta$ being uniformly distributed in $[0,2\pi]$.  
The numerical LEs agree well with the theoretical prediction in Eq.(\ref{LE})[see Fig.\ref{Fig1}(a)]. 
Besides the LE, we also calculate inverse of the participation ratio (IPR) and the fractal dimension (FD). 
For a normalized single-particle state, the IPR is defined by $P=\sum_j|\phi_j|^4$. 
In general, the IPR $P\propto L^{-\alpha}$, where $\alpha$ is the FD. 
For an extended state, $P\propto 1/L$ and $\alpha=1$, whereas the IPR approaches $1$ and $\alpha=0$ for a localized state. 
In the middle, a state with $0<\alpha<1$ is critical and has a self-similarity. 
In Fig.\ref{Fig1}(b), we present typical and  numerically extracted FDs in the ($V, E$) plane, whose values are indicated by colors. 
When $V$ is small the system is in the extended phase, whereas in the localized phase for a large enough $V$. 
In the intermediate region, it is in the mixed phase, consisting of extended, localized, and critical states. 
Critical states, indicating localization phase transition, only exist on the line of SPME, which is precisely described by Eq.(\ref{MB}) [red dash line in Fig.\ref{Fig1}(b)]. 
In Fig.\ref{Fig1}(c) we show typical spacial distributions of states of a system in the mixed phase ($V=0.78$), which clearly show localization properties of states before, on, and after the line of SPME.
The exact expression of SPME in Eq.(\ref{MB}) is rather complicated. 
Here, we illustrate some major behaviors of the SPME. 
On one hand, in Fig.\ref{Fig1}(d) we show lines of SPMEs in the ($V, E$) plane for systems with a fixed $b$ but different $\delta$. 
When $\delta=0$, the exact expression of SPME reduces to $E=2(-V\pm t)/b$, corresponding to two parallel lines in the ($V, E$) plane. 
However, spectra only interact with the upper line. 
This is the case which has been investigated  in Ref.\cite{Ganeshan2015}. 
With a non-zero $\delta$, the line of SPME turns into a lobe-shaped curve. 
As $\delta$ increases, the lobe becomes shorter and the tip of it moves upwards. 
When $\delta=\pi/2$, the exact expression of SPME reduces to $b^2E^2/4+V^2/(1-b^2)=t^2$. 
The line in ($V, E$) plane is the right half of an ellipse, whose axes are determined by $b$. 
Please see Fig.\ref{Fig1}(b) about how the line of SPME interacts with spectra. 
As $\delta$ further increases from $\pi/2$, the lobe becomes longer and the tip of it moves further upwards. 
Lines of SPMEs for systems with relative phases $\delta$ and $\pi-\delta$ are symmetric with respect to $E=0$ in the ($V, E$) plane, since Eq.(\ref{MB}) is invariant when putting both $\delta\rightarrow\pi-\delta$ and $E\rightarrow-E$. 
When $\delta=\pi$, the exact expression of SPME reduces to $E=2(V\pm t)/b$, another two parallel lines in the ($V, E$) plane, and spectra only interact with the lower line. 
On the other hand, in Fig.\ref{Fig1}(e) we present lines of SPMEs in the ($V, E$) plane for systems with a fixed $\delta$ but different $b$. 
The larger $b$ is, the smaller the lobe is, accompanied by a little bit distortion of the lobe (relative-phase-dependent).

\section{IV. Exact SPMEs in related models}

\begin{figure}[tbp]
	\begin{center}
		\includegraphics[scale=1.2, bb=5 67 132 169]{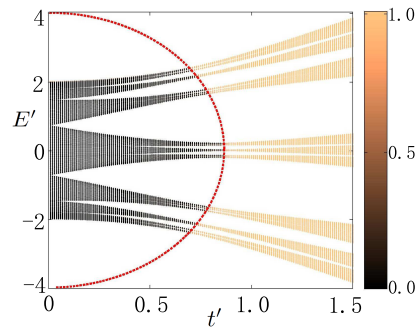}
		\caption{FDs of states, as function of the effective energy $E'=E_1+t_1e^{-p}\mathrm{sech}(p)\cos(\delta')$ and hopping amplitude $t'=\frac{t_1}{2}\mathrm{sech}(p)\sqrt{\tanh^2(p)\cos^2(\delta')+\sin^2(\delta')}$, for dual systems [Eq.(\ref{Dham})] with $V_1=1$, $p=\mathrm{arcosh}(2)\simeq 1.34$, $\delta'=\pi/2$, and $L=610$. 
		Red dash line corresponds to the theoretical SPME, whose exact expression is the same as Eq.(\ref{MB}) but with simultaneous replacements $t\rightarrow V_1$, $E\rightarrow E'$, and $V\rightarrow t'$.}
		\label{Fig2}
	\end{center}
\end{figure}

It is well-known that the classic AAH model ($b=0$) is self-dual. 
And recently in Ref.\cite{Wang2021} a dual relation has been established between the model in Eq.(\ref{Ham}) with $\delta=0$ and a generalized AAH model with long-range hoppings, which are two famous generalized AAH models hosting exact SPMEs \cite{Ganeshan2015,Biddle2010}. 
Two corresponding single-particle states of dual models have opposite localization properties: extended vs. localized, or both critical on the line of SPME. 
Here, with a general $\delta$, dual model of the Hamiltonian Eq.(\ref{Ham}) is given by
\begin{eqnarray}
H^D=&&t_1\sum_k\left[\sum_{l>0}e^{-pl+i\delta'}a^\dagger_ka_{k+l}+\sum_{l>0}e^{-pl-i\delta'}a^\dagger_ka_{k-l}\right]\notag\\
&&+2V_1\sum_k\cos(2\pi\beta k)a^\dagger_ka_{k},
\label{Dham}
\end{eqnarray}
where $p>0$ characterizes the long-range hopping. 
Introducing the duality transformation $a_k=\frac{1}{\sqrt{L}}\sum_jc_je^{i2\pi\beta kj}$, one can easily turn Eq.(\ref{Dham}) into the form of Eq.(\ref{Ham}) by summing geometric series. 
Relations among parameters of two models are given by
\begin{eqnarray}
t=V_1&&,\quad b=\mathrm{sech}(p),\quad E=E_1+t_1e^{-p}\mathrm{sech}(p)\cos(\delta'),\notag\\
V&&=\frac{t_1}{2}\mathrm{sech}(p)\sqrt{\tanh^2(p)\cos^2(\delta')+\sin^2(\delta')},\notag\\
&&\delta=\arccos\frac{\tanh(p)\cos(\delta')}{\sqrt{\tanh^2(p)\cos^2(\delta')+\sin^2(\delta')}}.
\end{eqnarray}
The last equation in the first line accounts for the energy shift between Hamiltonians, where $E_1$ is the energy of single-particle states of the Hamiltonian in Eq.(\ref{Dham}). 
Substituting above equations into Eq.(\ref{MB}), one can obtain the exact expression of SPME for the dual model, which is even more complicated. 
In Fig.\ref{Fig2} we present numerical FDs of single-particle states in the ($t', E'$) plane, for the system which is dual to the one shown in Fig.\ref{Fig1}(b). 
$E'=E_1+t_1e^{-p}\mathrm{sech}(p)\cos(\delta')$ is the effective energy, $t'=\frac{t_1}{2}\mathrm{sech}(p)\sqrt{\tanh^2(p)\cos^2(\delta')+\sin^2(\delta')}$ is the effective hopping amplitude, and $V_1$ sets the unit of energy ($V_1=1$). 
Spectra in Fig.\ref{Fig2} and Fig.\ref{Fig1}(b) are identical, but FDs show opposite localization properties, illustrating the dual relation between two models. 
The red dash line is the line of theoretical SPME, where critical states exist.

\begin{figure}[tbp]
	\begin{center}
		\includegraphics[scale=1.2, bb=73 60 199 163]{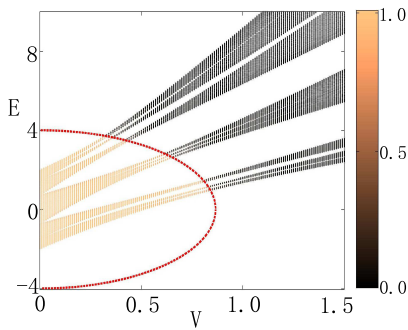}
		\caption{FDs in the ($V,E$) plane, for generalized AAH models with the potential in Eq.(\ref{ExtAAH}). 
		We choose $b=0.5$, $\delta=\pi/2$, $W=2$, and $L=610$. 
		Red dash line corresponds to Eq.(\ref{MB}).}
		\label{Fig3}
	\end{center}
\end{figure}

At last, we point out that there are generalized AAH models, whose SPMEs are also exactly described by Eq.(\ref{MB}). 
These models are defined by the Hamiltonian in Eq.(\ref{Ham}), but with the following on-site potential
\begin{equation}
V_j=2V\frac{W+\mathrm{cos}(2\pi\beta j+\theta+\delta)}{1-b\mathrm{cos}(2\pi\beta j+\theta)}.
\label{ExtAAH}
\end{equation}
Compared to Eq.(\ref{QP}), an additional dimensionless parameter $W$ is introduced in the numerator. 
When $W\rightarrow\infty$, $V\rightarrow0$, but $VW$ is finite, the cosine term in numerator disappears and the potential reduces to the one studied in Ref.\cite{Liu2021}. 
In this case the exact expression of SPME is $|E|=2t/b$, got from Eq.(\ref{MB}) by sending $V\rightarrow0$, while in Ref.\cite{Liu2021} it was obtained by a self-duality mapping. 
For the generalized AAH model with the potential in Eq.(\ref{ExtAAH}), in Fig.\ref{Fig3} we present examples of FDs in the ($V, E$) plane. 
Red dash line corresponds to Eq.(\ref{MB}), which precisely describes the SPME. 
Analytical computation of the exact expression of SPME can be easily done by following the same procedure shown above. 
In spite of causing no changes in the exact expression of SPME, the parameter $W$ changes spectra. 
The major effect of it is tilting spectra in the ($V, E$) plane [see Fig.\ref{Fig3} and Fig.\ref{Fig1}(b)]. 
For a positive (negative) $W$, the center of spectrum tilts upwards (downwards) as $V$ increases. 
Numerical data show that the larger $W$ is, the stronger the tilt is. 
The change of spectra alters the intersection between spectra and the line of SPME, and systems have different localization properties.

\section{V. Conclusion and discussion}

In this paper, we demonstrate the existence of exact SPMEs in a family of generalized AAH models, which are formed by introducing a relative phase into the model proposed by Ganeshan \cite{Ganeshan2015}. 
We analytically compute the LEs (or inverse of localization lengths) of all single-particle states and determine the exact expression of SPME, which are verified by numerical simulations. 
A physical perception of the rather complicated expression of SPME is also presented. 
We further demonstrate that the exact expression of SPME works for an even broad class of generalized AAH models. 
Moreover, we show that the exact SPME is related to the one in the dual AAH model, which has long-range hoppings and opposite localization properties. 
The generalized AAH model in the absence of relative phase has been realized in ultracold atoms, by using synthetic lattices of laser-coupled atomic momentum modes \cite{An2021}. 
The presence of SPME, not the exact expression of it, has been demonstrated, by showing that the localization order of states with the lowest and highest energies changes when the sign of b changes (or $\delta=0\rightarrow\delta=\pi$). 
Incorporating the relative phase and detecting signs of the change of SPME with respect to it are highly accessible in ultracold atoms. 
However, determining the exact position of SPME and studying its evolution with respect to the relative phase require more precise experimental controllability, or even new schemes.

\section{Acknowledgments}

This work is supported by the National Natural Science Foundation of China under grant No. 12134015 and No. 12175290.

\end{document}